\documentclass[fleqn,usenatbib]{mnras}

\usepackage{newtxtext,newtxmath}

\usepackage[T1]{fontenc}
\usepackage{ae,aecompl}

\usepackage{graphicx}
\usepackage{amsmath}
\usepackage{amssymb}

\usepackage[usenames,dvipsnames]{color}

\newcommand{\CI}{C\,\textsc{i}}

\newcommand{\FeII}{Fe\,\textsc{ii}}

\newcommand{\HI}{\textrm{H}\,\textsc{i}}

\newcommand{\Lya}{Ly$\alpha$}

\newcommand{\NHI}{$N(\textrm{H}\,\textsc{i})$}

\title[A novel limit on quantum foam]{A limit on Planck-scale froth with ESPRESSO\thanks{Based on observations collected at the European Organisation for Astronomical Research in the Southern Hemisphere, Chile [VLT program ID: 60.A-9508(A)]}}

\author[R. Cooke et al.]{
Ryan Cooke,$^{1,\thanks{E-mail: ryan.j.cooke@durham.ac.uk}}$
Louise Welsh,$^{1}$
Michele Fumagalli$^{1,2,3}$ and
Max Pettini$^{4}$
\\
$^{1}$Centre for Extragalactic Astronomy, Durham University, South Road, Durham DH1 3LE, UK \\
$^{2}$Institute for Computational Cosmology, Durham University, South Road, Durham DH1 3LE, UK \\
$^{3}$Dipartimento di Fisica G. Occhialini, Universit\`a degli Studi di Milano Bicocca, Piazza della Scienza 3, 20126 Milano, Italy\\
$^{4}$Institute of Astronomy, University of Cambridge, Madingley Road, Cambridge CB3 0HA, UK}

\date{Accepted XXX. Received YYY; in original form ZZZ}

\pubyear{2019}

\begin{document}
\label{firstpage}
\pagerange{\pageref{firstpage}--\pageref{lastpage}}
\maketitle

\begin{abstract}
Some models of quantum gravity predict that the very structure of spacetime is `frothy' due to quantum fluctuations. Although the effect is expected to be tiny, if these spacetime fluctuations grow over a large distance, the initial state of a photon, such as its energy, is gradually washed out as the photon propagates. Thus, in these models, even the most monochromatic light source would gradually disperse in energy due to spacetime fluctuations over large distances. In this paper, we use science verification observations obtained with ESPRESSO at the Very Large Telescope to place a novel bound on the growth of spacetime fluctuations. To achieve this, we directly measure the width of a narrow \FeII\ absorption line produced by a quiescent gas cloud at redshift $z\simeq2.34$, corresponding to a comoving distance of $\simeq5.8$~Gpc. Using a heuristic model where the energy fluctuations grow as $\sigma_{\rm E}/E=(E/E_{\rm P})^{\alpha}$, where $E_{\rm P}\simeq1.22\times10^{28}~{\rm eV}$ is the Planck energy, we rule out models with $\alpha\le0.634$, including models where the quantum fluctuations grow as a random walk process ($\alpha=0.5$). Finally, we present a new formalism where the uncertainty accrued at discrete spacetime steps is drawn from a continuous distribution. We conclude, if photons take discrete steps through spacetime and accumulate Planck-scale uncertainties at each step, then our ESPRESSO observations require that the step size must be at least $\gtrsim10^{13.2}l_{\rm P}$, where $l_{\rm P}$ is the Planck length.
\end{abstract}

\begin{keywords}
cosmology: theory --
elementary particles --
gravitation --
line: profiles --
quasars: absorption lines
\end{keywords}

\section{The Planck Scale}
\label{sec:intro}

At microscopic distance scales comparable to the Planck length,
$l_{\rm P}=\sqrt{\hbar G/c^{3}}\simeq1.62\times10^{-35}\,{\rm m}$,
it is thought that spacetime itself is subject to quantum fluctuations. If true, spacetime should appear ``fuzzy'' or ``frothy'', an effect that was termed ``quantum foam'' (also referred to as spacetime foam) by \citet{Whe63}. A foamy spacetime would cause minute uncertainties in the propagation of waves, such as the distance traversed by a photon, or its energy. If found, it would demonstrate that the nature of spacetime is probabilistic, rather than deterministic, and provide strong clues towards finding a unified description of gravity and quantum mechanics \citep[for an overview, see][]{Ame13}.

A variety of cosmological experiments have been conducted to place limits on models of quantum gravity. The most stringent constraint currently available is based on timing observations of distant $\gamma$-ray bursts (GRBs; \citealt{Abd09,Vas13}), which can be used to limit the in-vacuo dispersion of photons. Some models of quantum gravity predict that the photon speed depends on its energy \citep{Ame98,Mat05,JacLibMat06,KosMew08}, with the highest energy photons being the most affected. Over the immense cosmological distances to high redshift GRBs, the tiny shift of the propagation speed accumulates, and may produce a detectable difference in the arrival times of photons of different energy. The current GRB data disfavour quantum gravity theories that predict a variable speed of light at length scales, $l<l_{\rm P}/1.2$ \citep{Abd09}.

Currently, the most popular astrophysical probes of spacetime `fuzziness' are the phase delay and spatial blurring of cosmological sources. The first of these approaches was proposed by \citet{LieHil03}, building off a related proposal to search for quantum foam using gravitational wave interferometers \citep{Ame99}. The idea is relatively simple: some quantum gravity models predict that the period and wavelength of monochromatic photons gradually disperse as the wave propagates due to Planck-scale uncertainties. As the waveform travels further, its period and wavelength will increasingly deviate from the initial (i.e. emitted) values. When the waveform enters the aperture of a telescope (or interferometer), it will no longer represent a plane wave that uniformly illuminates the telescope aperture. As a result, if quantum foam scrambles the wavefront, an Airy disk diffraction pattern will not appear at the focus.

\citet{LieHil03} used the observation of an Airy disk in an image of  PKS\,1413+135 (at a distance of 1.2~Gpc), to suggest that first order fluctuations down to the Planck scale are ruled out. Shortly after their study, \citet{RagValMar03} extended this idea, and proposed that high-redshift cosmological point sources should experience spatial blurring due to the effects of quantum foam, and placed comparably strong limits on Planck-scale phenomena. However, the above results were contested soon after by \citet{NgChrvan03}, who highlighted that the cumulative effects of quantum foam depend on the choice of quantum gravity model (see also, \citealt{Cou03}).

Subsequent studies of spatial blurring \citep{ChrNgvan06,Ste07,Tam11,Chr11,Per11,Ste15} have narrowed the allowed model space by employing sources that emit high energy photons and are at larger cosmological distances. More recently, \citet{Per15} pointed out that when the distortions due to quantum foam become comparable to the wavelength of the photon, the intensity of a source decays to the point that the very \emph{detection} of a high redshift cosmological source places a strong bound on models of quantum gravity. Their work offers the tightest constraints yet, effectively ruling out the holographic model (see Section~\ref{sec:model}).

In this work, we present a novel limit on spacetime
foam using the widths of narrow spectral lines that
are seen in absorption against the light of a more
distant background quasar. In Section~\ref{sec:model},
we present an overview of the cosmological fluctuation
models that appear in the literature. We then extend this formalism to generalise the step size and the amount of uncertainty accumulated at each step. In Section~\ref{sec:obs},
we report a new bound on spacetime foam using the
\Lya\ forest, before offering an improved bound
using state-of-the-art observations with the Echelle SPectrograph for Rocky Exoplanets and Stable Spectroscopic Observations (ESPRESSO)
instrument \citep{Pepe10} on the Very Large Telescope.
We describe the future opportunities of this approach
in Section~\ref{sec:disc} before summarizing our main
conclusions in Section~\ref{sec:conc}. Throughout, we
assume a flat cosmology with
$H_{0}=67.8~{\rm km~s}^{-1}~{\rm Mpc}^{-1}$,
$\Omega_{\rm B,0}=0.04825$, and
$\Omega_{\rm M,0}=0.307$
\citep{Pla18}.

\section{The cosmological fluctuation model}
\label{sec:model}

Consider a precisely monochromatic laser placed at a cosmological distance. As the photons propagate toward our telescope, some models of quantum gravity predict that the photons gradually disperse in energy, leading to a measurable energy width of the photon ensemble. In this section, we derive the observed energy width of a cosmological laser.

\subsection{Literature Models}
\label{sec:litmodels}

The fundamental idea behind the following model is that spacetime
fluctuations accumulate over large cosmological distances. Most
of the studies described in Section~\ref{sec:intro} adopt the
following one parameter model to describe the
uncertainty ($\sigma_{l}$) of a distance measurement ($l$),
relative to the Planck length: $\sigma_{l}/l=(l_{\rm P}/l)^{\alpha}$,
where $\alpha$ describes how the quantum fluctuations grow as a
wave propagates (herein, we refer to $\alpha$ as the `growth factor'; note $0.5\le\alpha\le1.0$). One can show that this parameterisation leads to
similar uncertainties on the energy
\citep{Ngvan00,LieHil03}, $\sigma_{\rm E}/E=(E/E_{\rm P})^{\alpha}$,
where $E_{\rm P}=\hbar c/l_{\rm P}\simeq1.22\times10^{28}\,{\rm eV}$
is the Planck energy.

There are two proposals for how the accumulation of Planck-scale effects should grow over distance \citep[see][and papers thereafter by these groups]{LieHil03,NgChrvan03}.
According to \citet{NgChrvan03}, Planck-scale effects accumulate once every wavelength. Therefore, the fluctuations over distance, $L$, grow by a multiplicative factor
$C_{\alpha}=N^{1-\alpha}$, where $N=L/\lambda$ is the integer number
of wavelengths traversed by a photon (i.e. the number of times that quantum foam affects the photon).
In this case, the fluctuation model becomes:
$\sigma_{l}/l=(l_{\rm P}/l)^{\alpha}\,C_{\alpha}$. In an expanding Universe, we have:
\begin{equation}
N = \int \frac{{\rm d}r}{\lambda(z)} = \frac{c}{H_{0}\,\lambda_{0}} \int_{0}^{z} \frac{{\rm d}z}{E(z)} = \frac{D_{\rm C}}{\lambda_{0}}
\end{equation}
where $r$ is the proper distance, $\lambda_{0}$ is the observed wavelength of a photon emitted at redshift $z$, and $D_{\rm C}$ is the comoving distance. Thus, if $C_{\alpha}=N^{1-\alpha}$, a collection of precisely monochromatic photons emitted at redshift $z$ with wavelength $\lambda_{\rm em}$ would gradually disperse in energy, leading to an observed energy width (or, equivalently, a velocity width, $\Delta v_{\rm q}$):
\begin{equation}
\label{eqn:growth}
\sigma_{E}/E\equiv\Delta\lambda/\lambda\equiv\Delta v_{\rm q}/c=\frac{(l_{\rm P}/\lambda_{\rm em})^{\alpha}~(D_{\rm C}/\lambda_{\rm em})^{1-\alpha}}{1+z}
\end{equation}
It follows that one of the predictions of this formalism is that the relative widths of any two atomic transitions intrinsically dispersed by quantum foam should vary as:
\begin{equation}
\label{eqn:veleqn}
    \Delta v_{\rm q,1}/\Delta v_{\rm q,2} = \lambda_{2}/\lambda_{1}
\end{equation}
completely independent of: (1) the growth factor; (2) the distance travelled by the photon; and (3) cosmology.

Alternatively, Planck-scale effects may accumulate linearly with distance (i.e. $C_{\alpha}=N$); this is the original proposal put forward by \citet{LieHil03}. In this case, the observed velocity width (cf. Equation~\ref{eqn:growth}):
\begin{equation}
\label{eqn:growthLH}
\Delta v_{\rm q}/c=\frac{(l_{\rm P}/\lambda_{\rm em})^{\alpha}~(D_{\rm C}/\lambda_{\rm em})}{(1+z)^{1+\alpha}}
\end{equation}

In this work, we report limits on $\alpha$ based on the former approach (i.e. Equation~\ref{eqn:growth}), since it offers a more conservative limit on the growth of fluctuations. It is important to note that the above heuristic formalism, however plausible, has not yet been demonstrated by a model of spacetime quantization. Nevertheless, phenomenological studies, such as the one presented here, offer an opportunity to experimentally search for potential quantum effects on the structure of spacetime.

Since there is not a universally accepted theory
of quantum gravity, different spacetime foam models in the
family described above are
parameterised by different values of the
growth factor, $\alpha$.
The three most prominent
models of quantum foam that are discussed in the
literature are:

\begin{itemize}
    \item The random walk model \citep{DioLuk89,Ame99}, whereby the growth of quantum fluctuations is modelled as a random walk process, and is parameterised by $\alpha=1/2.$
    \item The holographic model of fluctuations \citep{Ngvan94} which is motivated by the holographic principle \citep{tHo93,Sus95}. In this model, the information in a three-dimensional volume can be encoded on a two-dimensional surface, resulting in $\alpha=2/3$.
    \item The so-called `standard' version of quantum foam, with $\alpha=1$, corresponding to the original proposal by \citet{Whe63}. In this model, the quantum fluctuations are not accumulated over distance, and there is therefore no obvious benefit in appealing to cosmological sources to test this model.
\end{itemize}

Note, the fluctuation formalism discussed above requires that $\alpha$ is in the range $0.5\le\alpha\le1$. In closing this section, we note that the derivation of the accumulation factor, $C_{\alpha}$, makes the following two assumptions: (1) uncertainties are accumulated at each step, where the step size is assumed to be equal to the wavelength; and (2) each step accumulates an uncertainty of $\pm l_{\rm P}(\lambda/l_{\rm P})^{1-\alpha}$ (i.e. only two possibilities), with equal probability. In the following subsection, we describe an alternative approach that makes different assumptions about the step size and the accumulation of uncertainty at each step.

\subsection{Convolutional fluctuation model}
\label{sec:convfluc}

We now describe a simple extension to the above formalism that allows us to generalise the step size taken by a particle, and the amount of uncertainty that is accumulated at each step. This simple model is primarily motivated by the aforementioned assumption that the uncertainty accumulated at each step is drawn from a two-point distribution, instead of a continuous distribution. One would naively expect that the two-point assumption exacerbates the effects of quantum foam. Instead, a more conservative approach is that the accumulated uncertainty is drawn from a continuous distribution.

Suppose a particle makes a step of size $\lambda$, and accumulates an uncertainty that is drawn from a uniform distribution between $\pm l_{\rm P}/2$. We are interested in deriving the dispersion accumulated over a distance $L$. We note that the above variables, $\lambda$ and $l_{\rm P}$, may be interpreted by the reader as the wavelength and Planck length, respectively. However, this is just for ease of comparison with previous work, and we stress that the following formalism is general to different choices of these length scales.

If at each step, the uncertainty of the step size is independent of the previous step, we can model this process as a rectangle (i.e. `tophat') function of width $l_{\rm P}$, repeatedly convolved with itself at each step. Assuming that the step size $\lambda\gg l_{\rm P}$, then the wave will take $N=L/\lambda$ steps, corresponding to $N$ convolutions of a rectangle function with itself. The resulting form of the spacetime foam-induced broadening function, $\Phi_{\rm q}$, can be calculated with Fourier transforms, assuming $N\gg1$,
\begin{eqnarray}
\left.\begin{aligned}
\Phi_{\rm q}&={\cal F}^{-1}[{\cal F}[{\rm rect}(l/l_{\rm P})]^{N}]\\
\Phi_{\rm q} &\simeq {\cal F}^{-1}[(1 - (\pi\,l_{\rm P}\,x)^2/6)^{N}]\\
\Phi_{\rm q} &\simeq {\cal F}^{-1}[\exp(-N\,(\pi\,l_{\rm P}\,x)^2/6)]
\end{aligned}\right.
\end{eqnarray}
By performing the inverse transform we conclude that the repeated convolution of an initially monochromatic wave that propagates with a uniform uncertainty between $\pm l_{\rm P}/2$, is a Gaussian of width $\sigma_{\rm L} = a_{0}\,l_{\rm P}\,\sqrt{L/\lambda}$, where $a_{0}=1/\sqrt{12}$ for a rectangle step probability.\footnote{We did not have to choose a rectangle function for the step probability; provided that the uncertainty associated with each step is independent of the previous step, the central limit theorem ensures that the only change to this functional form is to the value of the coefficient, $a_{0}$. For example, a Gaussian step probability of width $\sigma=l_{\rm P}$ will result in the same functional form, but with $a_{0}=1$.} If we adopt a step size that corresponds to the wavelength of the wave, as assumed in Section~\ref{sec:litmodels}, the accumulated fuzziness is diminished by a factor of $\sqrt{l_{\rm P}/\lambda}$ (i.e. $\sim14$ orders of magnitude for far-ultraviolet light) in the case of the $\alpha=0.5$ model. In other words, by allowing the accumulated fuzziness to be drawn from a uniform distribution rather than a two-point distribution with values $\pm l_{\rm P}(\lambda/l_{\rm P})^{1-\alpha}$, the effects of quantum foam are considerably diminished.

Finally, we note that the above formalism is not valid when the step size is of order $\sim l_{\rm P}$; this is satisfactory, since the accumulated effects of quantum foam would be \emph{substantial} in this case. Therefore, in what follows, we have chosen to model the step size $\lambda=\beta\,l_{\rm P}$. In this case, the accumulated energy spread of an initially monochromatic beam is given by:
\begin{equation}
    \label{eqn:growthCFM}
\Delta v_{\rm q}/c=\frac{a_{0}\,\sqrt{l_{\rm P}\,D_{\rm C}/\beta\,}}{\lambda_{\rm em}\,(1+z)}
\end{equation}
Note that the relative velocity width of any two atomic transitions, in this case, is identical to that described by Equation~\ref{eqn:veleqn}.

\section{Bounds on spacetime foam}
\label{sec:obs}

Transitions between atomic and molecular energy levels offer the best approximation to a monochromatic laser at high redshift. For the purposes of this work, we will use narrow absorption lines that are imprinted on the spectrum of a background quasar. This absorption line technique offers a fine, one-dimensional view of the gas that lies between us and the quasar. The atoms residing in the intervening gas cloud absorb the quasar light at discrete values corresponding to the energy levels of the atoms. The velocity width of the observed absorption lines include contributions from natural broadening (a Lorentzian function), as well as the turbulent and thermal motions of the atoms (a Maxwellian distribution).

An absorption line is therefore characterized by a Voigt profile with a Doppler width, $b$, which is simply related to the one-dimensional $1\sigma$ velocity interval along the line-of-sight, $\Delta v\equiv b/\sqrt{2}$. In the case of weak absorption lines, the natural broadening contribution to the line profile is undetectable, and only becomes apparent in the wings of the strongest absorption lines. In general, all atoms in a cloud experience the same amount of turbulent broadening ($b_{\rm t}$), while the thermal broadening ($b_{\rm th}$) depends on the gas kinetic temperature ($T_{\rm kin}$) and the atomic mass ($m$):
\begin{equation}
\label{eqn:therm}
b_{\rm th}^{2}=2\,k_{\rm B}\,T_{\rm kin}/m
\end{equation}
where $k_{\rm B}$ is the Boltzmann constant. The total Doppler width of an absorption line is then given by $b^{2}=b_{\rm t}^{2}+b_{\rm th}^{2}$. Therefore, the heaviest ions produce the narrowest absorption features; furthermore, absorption lines of the same ion that arise from the ground state are expected to exhibit the same broadening. Using this guidance, one could in principle identify quantum foam by measuring the relative widths of several absorption lines from a heavy atom, such as \FeII, and searching for a wavelength dependent velocity width of the form given by Equation~\ref{eqn:veleqn}.

There are also wavelength dependent contributions to the line broadening that one must consider due to the spectrograph that is used to record the data. The main contribution to the line broadening for a narrow absorption line is the instrument spectral resolution. For reference, the world's premier optical echelle spectrographs have a full-width at half-maximum (FWHM) spectral resolution as low as a few ${\rm km~s}^{-1}$.

Finally, we note that quantum foam would broaden an absorption line in a similar fashion to the instrument broadening function. Specifically, the intrinsic (Voigt) profile of the absorption line generated by a gas cloud would be convolved with a Gaussian of width $\Delta v_{\rm q}$ due to the cumulative effects of quantum foam. Then, as the light passes through the spectrograph, it would be further convolved by the instrument broadening function (usually approximated by a Gaussian).\footnote{In reality, the functional form of the instrument broadening function is not a Gaussian. The instrumental profile typically exhibits non-Gaussian wings and an asymmetry, both of which depend on wavelength and spectral order.} Thus, if quantum foam broadens the line profile, it will appear as an `effective' instrument FWHM profile of width $\sqrt{v_{\rm FWHM}^{2}+\Delta v_{\rm q}^{2}}$ (i.e. slightly broader than the actual instrumental FWHM).

In the presence of quantum foam, the equivalent widths of absorption lines would be preserved. It is therefore important to \emph{directly} resolve the width of a narrow absorption line to place a limit on quantum foam using this approach. In other words, a curve-of-growth analysis is insufficient to infer quantum foam-induced broadening. This is unfortunate, because the existence of intrinsically narrow (unresolved) spectral features has been inferred in several quasar absorption line systems \citep[see e.g.][]{Jor09,JorWolPro10}. We must therefore resort to systems with simple kinematics, but with lines that are not too narrow compared to the instrument resolution. In the case of absorption line profiles that are marginally resolved, we need to accurately determine the instrumental FWHM.

We now consider two lines of evidence that place strong limits on the accumulation properties of quantum foam.

\begin{figure}
	\includegraphics[width=\columnwidth]{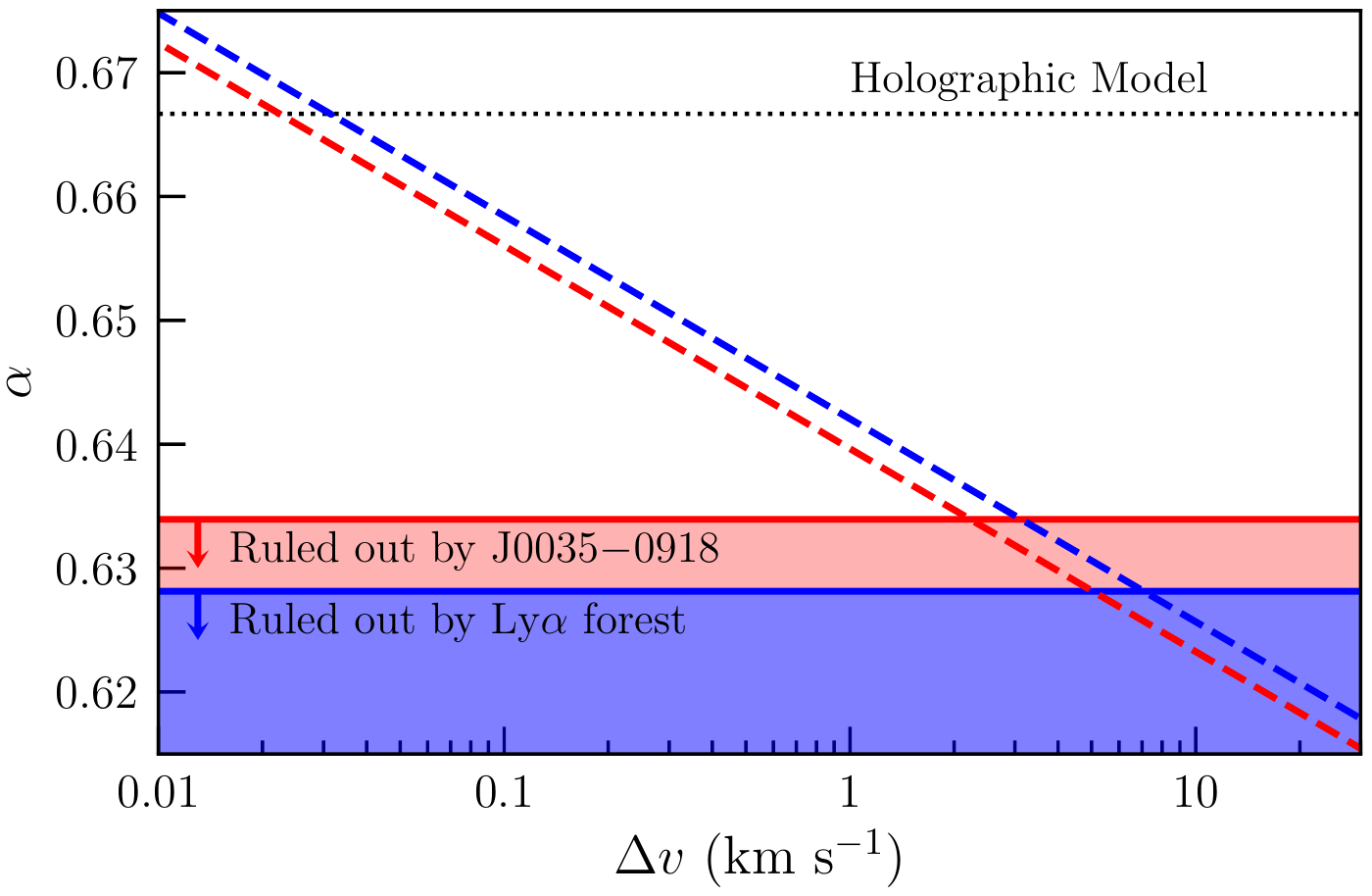}
\caption{The dashed diagonal lines show how the velocity width of a spectral line depends on the accumulated effects of quantum foam, parameterized by the growth factor, $\alpha$. The red dashed line represents an \FeII\,$\lambda1608$\AA\ absorber at redshift $z=2.34$ (i.e. the case of J0035$-$0918), while the blue dashed line represents a \Lya\ absorber (\HI\,$\lambda1215$\,\AA) at redshift $z=2.5$. The blue and red shaded bands indicate the range of $\alpha$ models that are ruled out by the measured widths of the \Lya\ forest and J0035$-$0918, respectively. The black horizontal dotted line labelled `Holographic Model' indicates a model with $\alpha=2/3$.
    }
    \label{fig:alpha}
\end{figure}

\subsection{\Lya\ forest absorption}

First, we consider the multitude of absorption lines that comprise the \HI\ \Lya\ forest \citep{Lyn71}. Cosmological hydrodynamic simulations of the \Lya\ forest indicate that these gas clouds predominantly arise from low density gas in the intergalactic medium \citep[see][and references therein]{Mei09}. Observations of either multiply-imaged or multiple nearby quasar sightlines \citep{Sme92,Bec94,Sme95,Fan96,DOd06} indicate that these lines of sight predominantly intersect large scale structures whose motions mainly contain contributions from the Hubble flow \citep{Rau05} and the thermal motions of the intersected gas. Since the broadening of the \Lya\ forest lines is dominated by both the thermal motions of the gas and Hubble broadening, we adopt the conservative assumption that the minimum velocity width of the lines comprising the \Lya\ forest provides an upper limit on the possible broadening due to quantum foam, and a corresponding lower limit on the growth factor.

The \Lya\ forest absorption lines have been found to exhibit Doppler widths as low as $\sim10~{\rm km~s}^{-1}$ at redshift $z\simeq2.5$ \citep{RudStePet12}. Such absorption features are well-resolved by current echelle spectrographs, which leads to a direct measure of the line width. Given the assumed cosmology \citep{Pla18}, the comoving distance to these \HI\ \Lya\ absorbers ($\lambda_{\rm em}=1216$\AA) is $D_{\rm C}\simeq6~{\rm Gpc}$, leading to a bound on the growth factor, $\alpha\ge0.628$. This limit, together with the growth curve (Equation~\ref{eqn:growth}), is presented in Figure~\ref{fig:alpha} (blue solid and dashed lines, respectively).

The \Lya\ forest therefore rules out quantum gravity models that require a growth factor of $\alpha=0.5$, including the random walk model. Said differently, if quantum fluctuations accumulate over distance with $\alpha=0.5$, the velocity width of \Lya\ absorbers at $z\simeq3$ would be $\Delta v_{\rm q}\simeq 5.6\times10^{8}~{\rm km~s}^{-1}$ (refer to Equation~\ref{eqn:growth}).\footnote{Such a large number results from the exponential terms in Equation~\ref{eqn:growth}, combined with the large numbers involved; a relative small change in $\alpha$ leads to a large change to the bound on the velocity width. Conversely, in order to make a significant improvement on the $\alpha$ limit, one requires a much stronger bound on the velocity width.} Thus, the very existence of absorption features associated with the \Lya\ forest is incompatible with models that predict a growth factor $\alpha=0.5$.

\begin{figure}
	\includegraphics[width=\columnwidth]{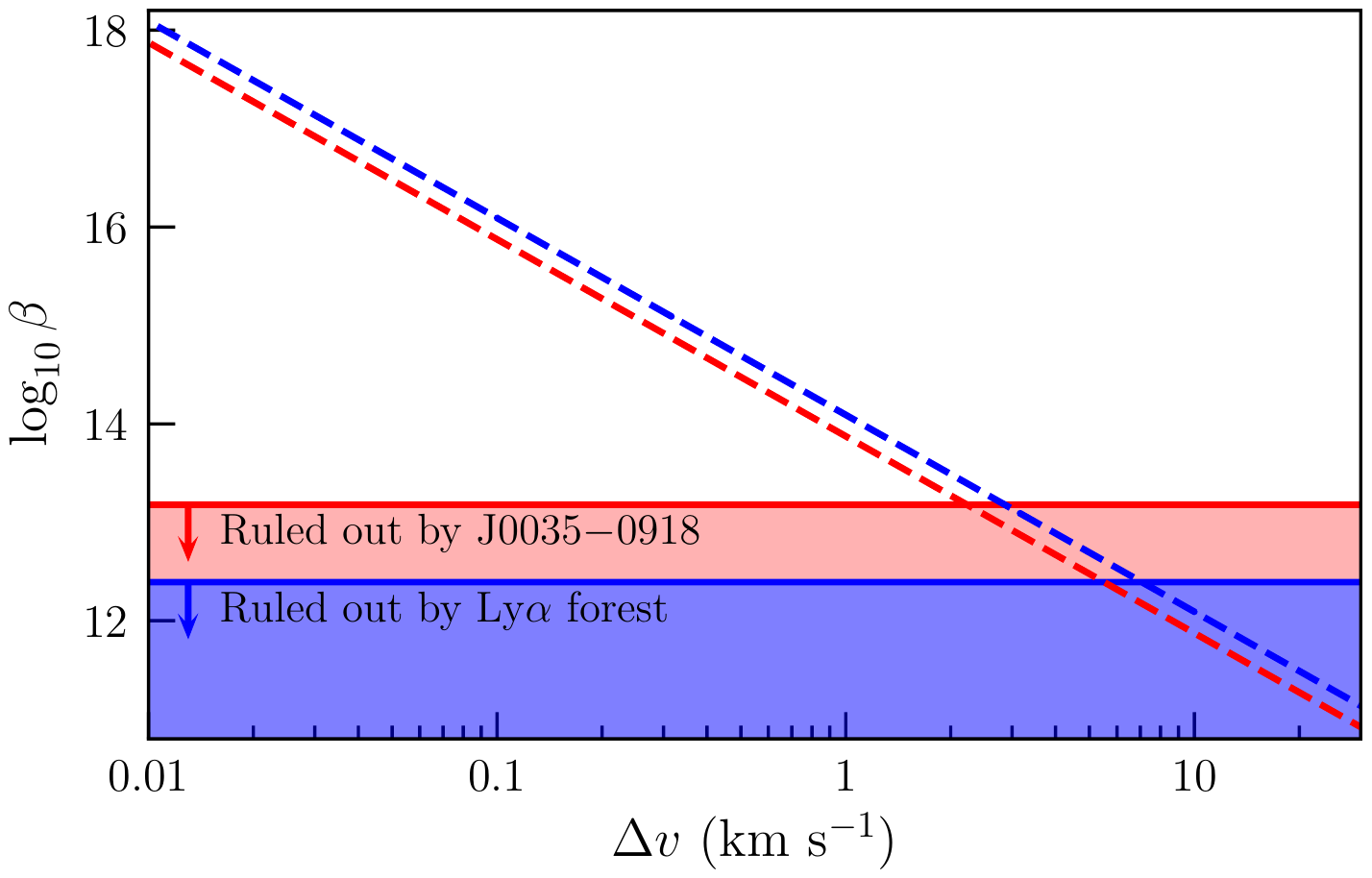}
\caption{Same as Figure~\ref{fig:alpha}, but shows the limits on the growth factor, $\beta$, for the convolution fluctuation model (see Section~\ref{sec:convfluc}). This model suggests that, if photons take discrete steps through spacetime and accumulate a Planck-scale uncertainty at each step, then the step size must be at least $\beta l_{\rm P}$.
    }
    \label{fig:beta}
\end{figure}

If we instead consider the convolutional fluctuation model proposed in Section~\ref{sec:convfluc}, the \Lya\ forest requires that $\beta\ge10^{12.4}$, implying that if photons accumulate a Planck-scale uncertainty at discrete steps in spacetime, then the step size must be $\beta\,l_{\rm P}\ge4\times10^{-23}\,{\rm m}$ to be consistent with the \Lya\ forest. The blue shaded band in Figure~\ref{fig:beta} indicates the values of $\beta$ that are ruled out by observations of the \Lya\ forest.

\subsection{Damped \Lya\ systems}

To place a tighter bound on the growth of quantum fluctuations, we need to identify absorption line systems that: (1) contain quiescent gas; (2) are decoupled from the Hubble flow; and (3) contain gas that is sufficiently cold to minimise the effects of thermal broadening. These properties are generally satisfied by damped \Lya\ systems (DLAs), which are clouds of mostly neutral gas intersected by lines of sight to unrelated, background quasars. DLAs are defined to have \HI\ column densities that exceed \NHI~$\ge10^{20.3}~{\rm cm}^{-2}$ \citep{Wol86}, and appear to be associated with a range of galaxy types (\citealt{Pon08,Fum15,Kro17}; for a general review of DLAs, see \citealt{WolGawPro05}). The metallicity distribution function of DLAs evolves slowly with redshift \citep{Raf12,Jor13}, and is peaked at a metallicity $\sim1/30$ of solar \citep{Pet97,Pro03,Raf12,Jor13}. The DLA population also displays a metallicity-velocity width relation \citep{Led06,Mur07,Pro08,Jor13,Coo15}, which is thought to be tied to an underlying mass-metallicity relation. Taken together, the above characteristics fall on a `fundamental plane' between metallicity, redshift, and velocity \citep{Nee13}; the DLAs with the simplest kinematics are those that have the lowest metallicity.

The most metal-poor DLAs are clouds of gas that have been enriched by at most a few generations of stars \citep{Wel19a}, and in some cases, exhibit just a single absorption component of kinematically quiescent gas \citep[see the example line profiles of the systems reported by][]{Pet08,Coo11a,Coo11b}. Of the metal-poor DLAs currently known, the most quiescent gas is exhibited by the absorber at $z_{\rm DLA}\simeq2.34$ toward the quasar J0035$-$0918, first reported by \citet{Coo11a}. \citet{Dut14} collected and analyzed new data of this system that covered several strong \FeII\ lines, thereby allowing the kinematics and chemistry of this gas cloud to be pinned down (see also, \citealt{Coo15}).

Bolstered by the quiescence of this metal-poor DLA, we requested $3\times2100\,{\rm s}$ exposures with ESPRESSO in 4UT mode during the science verification phase. We collected an exquisite, high resolution spectrum of the absorption lines associated with this DLA, allowing us to measure the detailed isotopic chemistry of the gas cloud and derive the gas kinematics \citep[see][for the detailed analysis of these new data, including the updated chemistry of this system]{Wel19b}. The final combined signal-to-noise ratio of the data near \FeII\,$\lambda1608$\AA\ is S/N~$\simeq20$ per wavelength bin. Even at the resolution of ESPRESSO in 4UT mode (FWHM velocity resolution of $4.28~{\rm km~s}^{-1}$), the absorption profile of this DLA is well-represented by a single, narrow component \citep[see][]{Wel19b}.

To model the absorption lines, we use the Absorption LIne Software (ALIS;\footnote{ALIS is publicly available from\\ \url{https://github.com/rcooke-ast/ALIS}} see \citealt{Coo14} for details about this software). To place a limit on the velocity width due to quantum foam, we need to first measure the widths of the absorption line profiles, and remove the instrumental contributions to the line widths. We account for the instrumental FWHM of the line profiles by measuring the widths of the ThAr calibration lines at the measured wavelengths of the DLA absorption features (see \citealt{Wel19b} for further details).

After accounting for the measured instrumental broadening, \citet{Wel19b} find that the line profiles are entirely dominated by thermal broadening, with a temperature $T_{\rm kin}=9100\pm500~{\rm K}$. The absorption line that is most sensitive to quantum foam-induced broadening is \FeII\,$\lambda1608$\,\AA, since this is the shortest wavelength line of the highest atomic mass element detected. The Doppler parameter of the \FeII\ absorption, $b_{\rm th}=2.70\pm0.15$, makes it the narrowest line directly measured with the available ESPRESSO data. We adopt the conservative assumption that this line width provides an upper limit on quantum foam-induced broadening. Using the relation $\Delta v_{\rm q}=b_{\rm th}/\sqrt{2}$, we place a $3\sigma$ lower limit on the allowed values of the growth factor $\alpha\ge0.634$, which is represented by the red shaded band in Figure~\ref{fig:alpha}. Using instead the convolutional fluctuation model described in Section~\ref{sec:convfluc}, our ESPRESSO observations require $\beta\ge10^{13.2}$ ($3\sigma$), leading to a limit on the step size, $\beta\,l_{\rm P}\ge2.4\times10^{-22}\,{\rm m}$ (see red shaded band in Figure~\ref{fig:beta}).

Our reported limit on the allowed range of $\alpha$ using J0035$-$0918 is almost as competitive as the best available limits using the image blurring of cosmological point sources ($\alpha\ge0.65$ \citealt{Chr11,Tam11,Per11}). The most competitive bound on the accumulation power of quantum foam is based on the mere detection of distant point sources at GeV energies ($\alpha\gtrsim0.72$ \citealt{Per15}). Nevertheless, the narrow spectral line observations that we present here offer a complementary and independent limit on quantum foam.

\subsection{Blazars}

For completeness, we also report a limit on $\beta$ using the \citet{Per15} approach, which is based on the detection of very high energy (VHE) sources. In this regime, when the accumulated uncertainty becomes comparable to the wavelength, the simple detection of a source can rule out models of quantum foam. The most distant cosmological source\footnote{For a list of VHE sources, see:\\ \url{http://tevcat.uchicago.edu/}} that has been detected in VHE emission, with a secure redshift, is the flat-spectrum radio quasar PKS~1441+25, at a redshift $z=0.939$ \citep{Ahn15,Abe15}. This source has been detected in gamma-ray emission >100~GeV, which corresponds to a limit $\beta\ge10^{23.3}$, or equivalently a step size of $\beta\,l_{\rm P}\ge3.8\times10^{-12}~{\rm m}\equiv3\times10^{5}\,\lambda$, where $\lambda$ is the wavelength of a 100~GeV photon. In other words, the effects of quantum foam --- if present --- are being accumulated over a distance that is at least 300,000 times larger than the wavelength of a 100~GeV photon.

\section{Future opportunities}
\label{sec:disc}

There are two obvious possibilities to improve upon the current limit using spectral line observations: (1) acquire higher spectral resolution observations of intrinsically narrow absorption line profiles; or (2) search for spectral lines or abrupt changes to the spectral shape at higher energies. We briefly explore each of these possibilities below.

The first possibility could be readily realised with dedicated observations of known \CI\ absorbers \citep[e.g.][]{JorWolPro10,Not18} or CO molecular absorption lines \citep[e.g.][]{Not11} towards high redshift quasars. Both \CI\ and CO molecular absorption lines probe the cold neutral medium of galaxies, where the thermal contribution to the line width is minimal. Most of the known \CI\ and CO absorbers are relatively metal-rich with tens of absorption components; the overall kinematics of these absorbers can exceed several hundred km~s$^{-1}$, but in some cases, the \emph{individual} \CI\ and CO lines are apparently not blended with other features. The highest spectral resolution observations with ESPRESSO in 1UT mode (${\rm FWHM}\simeq1.5~{\rm km~s}^{-1}$) might permit a measure of quantum foam-induced broadening down to a level of a few hundred ${\rm m~s}^{-1}$, provided that the amount of broadening due to the instrument is well-determined. Such a measure would deliver a limit $\beta\ge10^{15}$ or an equivalent limit on the growth factor of $\alpha\sim0.66$, which is approaching the value expected for the holographic model.

Looking forward, in order to be competitive with the \citet{Per15} approach, the best opportunity is to use very high energy sources at cosmological distances. To give an illustrative example, if the spectral shape changes by just $\sim10$ percent (i.e. $\Delta E/E\sim0.1$) at an energy of $\sim10~{\rm TeV}$, then a source at redshift $z\sim1$ would provide a limit $\alpha\gtrsim0.76$, or $\beta\gtrsim10^{30}$. Such an experiment may become possible with the Cherenkov Telescope Array, which is expected to be online in the next five years.

\section{Summary and Conclusions}
\label{sec:conc}

Using data acquired with the recently commissioned
ESPRESSO spectrograph at the European Southern Observatory Very Large Telescope, we have placed a
novel limit on the existence of spacetime foam
based on the intrinsic widths of measured spectral
lines. The main conclusions of this paper can
be summarized as follows:\\

\noindent ~~(i) We describe a novel approach to place
a bound on quantum foam-induced fluctuations, based on
the measured energy widths of rest-frame ultraviolet
absorption lines. We employ a commonly used heuristic model, which contains a single parameter $\alpha$ that characterises how the quantum fluctuations grow as a wave propagates.

\smallskip

\noindent ~~(ii) We also present a new formalism to model the accumulated energy fluctuations in the event that photons take discrete spacetime steps as they propagate. This simple model is characterised by a single parameter, $\beta$, where the step size is given by $\beta\,l_{\rm P}$, and $l_{\rm P}$ is the Planck length.

\smallskip

\noindent ~~(iii) We first apply the above models to observations
of the \Lya\ forest, which already place a strong limit
on the growth factors, $\alpha\ge0.628$ and $\beta\ge10^{12.4}$. Indeed, the very
existence of the \Lya\ forest
indicates that the random walk model (corresponding to $\alpha=0.5$)
is ruled out.

\smallskip

\noindent ~~(iv) Currently, our strongest bound on
spacetime foam-induced fluctuations is derived from the
narrow \FeII\,$\lambda1608$ absorption line of the
near-pristine DLA at $z_{\rm DLA}\simeq2.34$ towards
the quasar J0035$-$0918. Here, we report a conservative $3\sigma$ limit on the growth factors, $\alpha\ge0.634$ and $\beta\ge10^{13.2}$. The latter corresponds to a photon step size of $\beta\,l_{\rm P}\ge2.4\times10^{-22}\,{\rm m}$.

\smallskip

\noindent ~~(v) We suggest future opportunities to
use narrow spectral line observations to place more stringent
limits on models of spacetime foam. In the immediate future,
perhaps the best opportunity is to directly measure the widths of
\CI\ or CO absorption lines towards known metal-rich DLAs
with an ultra-high resolution echelle spectrograph, such
as ESPRESSO on the VLT. Such observations may just be able
to test the holographic model of spacetime foam, independent
of the phase and image blurring approaches that have previously been used.

\smallskip

\noindent ~~(vi) Previous work by \citet{Per15} has shown that gamma-ray detections of high redshift blazars offer the tightest current bounds on $\alpha$. Using this approach, we place a strong limit on the growth factor $\beta\ge10^{23.3}$, which is equivalent to a step size of $3\times10^{5}\,\lambda$, where $\lambda$ is the wavelength of a 100~GeV photon.

\smallskip

We find that our reported limits on quantum foam-induced
fluctuations are competitive with the blurring of
cosmological point sources, and provide complementary
evidence to support the current limits on $\alpha$ using
that approach. Owing to the extra sensitivity of the
fluctuation models at higher energy,
spectral line/shape observations at TeV energies may
further narrow the range of allowed models, and provide
strong clues towards finding a unified description of
gravity and quantum mechanics.

\section*{Acknowledgements}
We are grateful to the staff astronomers at the VLT for their assistance with the observations. During this work, R.~J.~C. was supported by a
Royal Society University Research Fellowship (UF150281).
R.~J.~C. acknowledges support from STFC (ST/L00075X/1, ST/P000541/1).
This project has received funding from the European Research Council 
(ERC) under the European Union's Horizon 2020 research and innovation 
programme (grant agreement No 757535).
This research has made use of NASA's Astrophysics Data System, and the following software packages:
Astropy \citep{Ast13},
Matplotlib \citep{Hun07},
and NumPy \citep{van11}.

\bsp
\label{lastpage}
\end{document}